\documentclass[traditabstract]{aa}
\usepackage{txfonts}
\usepackage{natbib,amstext,textcomp,graphicx,microtype,xcolor}
\usepackage{enumitem,textcomp}
\bibpunct{(}{)}{;}{a}{}{,}

\newcommand{\sbs}[1]{_\text{#1}}  %subscript
\newcommand{\total}{\ensuremath{\text{d}}}
\newcommand{\micron}{\ensuremath{\text{\textmu{m}}}}

\begin{document}

%------------------------------------------------------------------

\title{Relating grain size distributions in circumstellar discs to the spectral index at millimetre wavelengths}
\institute{Astrophysikalisches Institut und Universit{\"a}tssternwarte,
           Friedrich-Schiller-Universit{\"a}t Jena,
           Schillerg{\"a}{\ss}chen 2--3, 07745 Jena, Germany, \email{torsten.loehne@uni-jena.de}
}
\author{T.~L\"ohne}

\abstract{The excess emission seen in spectral energy distributions (SEDs) is commonly used to infer the properties of the emitting circumstellar dust in protoplanetary and debris discs. Most notably, dust size distributions and details of the collision physics are derived from SED slopes at long wavelengths. This paper reviews the approximations that are commonly used and contrasts them with numerical results for the thermal emission. The inferred size distribution indexes $p$ are shown to be greater and more sensitive to the observed sub(mm) spectral indexes, $\alpha\sbs{mm}$, than previously considered. This effect results from aspects of the transition from small grains with volumetric absorption to bigger grains that absorb and emit near to their surface, controlled by both the real and the imaginary part of the refractive index. The steeper size distributions indicate stronger size-dependence of material strengths or impact velocities or, otherwise, less efficient transport or erosion processes. Strong uncertainties remain because of insufficient knowledge of the material composition, porosity, and optical properties at long wavelengths.
}

\maketitle

%===========================================================================
\section{Introduction}
%===========================================================================
The excess emission commonly observed in spectral energy distributions (SEDs) reflects the properties and the evolution of the emitting circumstellar dust. Material from a range of grain sizes, radial distances, grain compositions, etc. contributes to the observed flux density. In turn, the distributions across these grain properties result from the past and ongoing processes that create and remove the dust.

The shape of an observed disc-integrated SED is a convolution of several dust properties and processes. A prominent example for the degeneracies that occur is the small grains at greater distances that can have the same temperatures as bigger grains closer to the star, where big and small are relative to the stellar spectrum. In a similar way, the temperature distribution and the resulting SED can be broadened by both a range of grain sizes and a range of distances. A narrow ring containing small and big grains can mimic a wide ring with a narrow size distribution.

The analysis of long-wavelength emission suffers less from a temperature degeneracy. Grains are inefficient emitters in the Rayleigh regime, that is, at wavelengths much longer than their own scale. At long wavelengths, the contribution from smaller grains to the SED is small. With these warmer grains out of the equation, the range of temperatures involved is reduced; temperatures and grain size become less entangled. This allowed, for example, \citet{draine2006} to derive an analytic approximation that relates the spectral index in the mm regime, $\alpha\sbs{mm}$, to the power-law index of the underlying grain size distribution, $p$. The linear relation that results is:
\begin{equation}
  p = \frac{\alpha\sbs{mm} - \alpha\sbs{Pl}}{\beta\sbs{s}} + 3\label{eq:draine},
\end{equation}
where $\alpha\sbs{Pl}$ is the slope of the Planck function at the appropriate temperature and $\beta\sbs{p}$ is the power-law index of the wavelength-dependent opacity.

Knowledge of the size distribution index provides deep insights into the physical environment of the dust. In protoplanetary discs, the processes of agglomeration and compaction result in net grain growth and shape the size distribution \citep[e.\,g.][]{dominik+2007}. In debris discs, disruption and erosion steadily transfer mass from bigger to smaller objects in a collisional cascade \citep[e.\,g.][]{krivov2010,matthews+2014}. On the one hand, the resulting dust size distribution is steepened by the material strength decreasing with grain size \citep{o'brien+greenberg2003,wyatt+2011} and by successive collisional damping as material is moving down the collisional cascade towards smaller sizes \citep{pan+schlichting2012}. On the other hand, an interplay of radiation pressure and dynamical excitation \citep{thebault+wu2008} and transport processes such as Poynting--Robertson or stellar wind drag \citep{reidemeister+2011,wyatt+2011} flatten the size distribution. The canonical value of $p = 3.5$ is reached for constant impact velocities and material strength \citep{dohnanyi1969}. Plausible ranges for average size distribution slopes roughly cover $3 \lesssim  p \lesssim 4$. 

The small-size cutoff due to radiation pressure blowout induces a wavy deviation where slopes can exceed this range locally \citep{campo-bagatin+1994a}. That waviness partially transcends to the observed SEDs \citep{thebault+augereau2007}.
However, Eq.~(\ref{eq:draine}) is derived under certain assumptions, with alternative approximations available for other assumptions. The aim of this work is, therefore, to delimit the parameter ranges in which these approximations apply and when it is best to use a full numerical modelling and fitting of thermal emission, which can be considered the benchmark. 

Section~\ref{sec:Approximations} provides a review of the basic equations that govern all thermal emission models. Possible approximations are discussed, including those that lead to Eq.~(\ref{eq:draine}). Section~\ref{sec:Numerics} contains a comparison of numerical and analytic results, revealing and explaining discrepancies that significantly affect the desired direct link between observables and dust physics. The immediate impact on the analysis of literature data is discussed in Sub-section \ref{sec:Application}. Sections~\ref{sec:Discussion} and \ref{sec:Conclusions} discuss the implications more broadly and summarize the findings.

%===========================================================================
\section{Analytic approximations}\label{sec:Approximations}
%===========================================================================
% In this section, we recapitulate the analytic model and simplifying assumptions that allow to relate observed spectral energy distributions to underlying size distributions.

An optically thin circumstellar dust ring at a distance $d$ from the observer produces a total thermal flux density,
\begin{equation}
  F_\nu (\lambda) = \int_{s\sbs{min}}^{s\sbs{max}}\limits \pi B_\nu [T(s), \lambda] \, Q\sbs{abs}(T(s), s, \lambda) \, \frac{s^2}{d^2} \, N(s) \, \total s,\label{eq:basic}
\end{equation}
% \tl{discuss radial dependence?}
with $B_\nu$ the Planck function, $T(s)$ the grain Temperature, $\lambda$ the wavelength, $Q\sbs{abs}$ the absorption and emission efficiency, $N(s)$ the differential size distribution, and $s\sbs{min}$ and $s\sbs{max}$ the minimum and maximum grain radii.
In general, the absorption efficiency depends on wavelength, material, and temperature as well as grain size and shape. 
In thermal equilibrium, grain temperature $T$ and distance $r$ from the star are related through \citep{krivov+2008}:
\begin{equation}
  r = \frac{R_*}{2} \sqrt{\frac{\int C\sbs{abs}(T, s, \lambda) F_{\lambda,*} \total\lambda}{\int C\sbs{abs}(T, s, \lambda) B_\lambda(T, \lambda) \total\lambda}},\label{eq:temps}
\end{equation}
where $C\sbs{abs} \equiv \pi s^2 Q\sbs{abs}$ is the effective absorption cross-section per grain.
Whenever the temperature can be considered constant over the relevant range of grain sizes, Eq.~(\ref{eq:basic}) can be simplified to 
\begin{equation}
  F_\nu (\lambda) \approx \frac{B_\nu (T, \lambda)}{d^2} \, \int_{s\sbs{min}}^{s\sbs{max}}\limits C\sbs{abs}(T, s, \lambda) \, N(s) \, \total s\label{eq:Fnu}.
\end{equation}
Grain temperature is discussed in greater detail in Sect.~\ref{sec:temp}.

Defining further a total cross-section,
\begin{equation}
  C\sbs{abs,total}(\lambda) \equiv \int C\sbs{abs}(s, \lambda) N(s) \total s\label{eq:Cabstot}
\end{equation}
and disc mass
\begin{equation}
  M\sbs{d} = \int M(s) N(s) \total s,\label{eq:Mtot}
\end{equation}
where $M(s)$ is the mass of a single grain with radius $s$, the wavelength-dependent total opacity is given by
\begin{equation}
  \kappa(\lambda) \equiv \frac{C\sbs{abs,total}(\lambda)}{M\sbs{d}}.\label{eq:kappa-def}
\end{equation}
After inserting Equations (\ref{eq:Cabstot})--(\ref{eq:kappa-def}) into (\ref{eq:Fnu}), the resulting flux density is
\begin{equation}
  F_\nu (\lambda) \approx \frac{B_\nu (T, \lambda)}{d^2} \, \kappa(\lambda) M\sbs{d}.\label{eq:Fnu-kappa}
\end{equation}
The measured flux density can thus be decomposed into the underlying blackbody and the opacity. If $\kappa$ follows a power law,
\begin{equation}
  \kappa \propto \lambda^{-\beta},\label{eq:kappa}
\end{equation}
the opacity index $\beta$ is related to the spectral index via
\begin{equation}
  \log \frac{F_\nu}{B_\nu} = -\beta\log\lambda + \text{const},
\end{equation}
which is equivalent to
\begin{equation}
  \underbrace{\frac{\log F_{\nu,2}/F_{\nu,1}}{\log\lambda_2/\lambda_1}}_{= -\alpha\sbs{mm}} - \underbrace{\frac{\log B_{\nu,2}/B_{\nu,1}}{\log\lambda_2/\lambda_1}}_{= -\alpha\sbs{Pl}} = -\beta\label{eq:gamma-beta}\end{equation}
for any given pair of wavelengths $\lambda_1$ and $\lambda_2$.
Assuming that the temperature is known and constant over the range of relevant grain sizes, the blackbody component can thus be removed from observed fluxes. The spectral index of the opacity, and hence the size distribution, can be derived \citep[e.\,g.][]{ricci+2012,ricci+2015b,macgregor+2016a,marshall+2017}.
% \tl{find more such papers}

\begin{figure}
  \centering
  \includegraphics{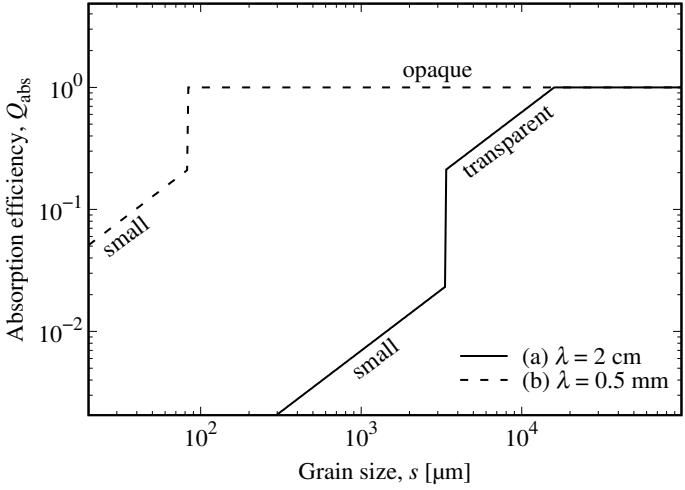}
  \caption{Schematic comparison of the two different cases (a) and (b) for the transition from small grains that absorb by volume to big, opaque grains that absorb by cross-section. Case (a) has an intermediate size interval where the grains are already big compared to the wavelength but still transparent. In case (b) the transition from small to opaque is direct. The values assumed in equations~(\ref{eq:Qabs,s2}), (\ref{eq:Qabs,t2}), and (\ref{eq:alpha}) are $\beta\sbs{s} = 1.6$, $A_0 = 0.63~\text{mm}^{-1}$, $\lambda_0 = 1$~mm, $n = 3$, $y = n^2$.\label{fig:s_c}}
\end{figure}

While the details of $C\sbs{abs}(s,\lambda)$ depend on grain morphology, structure, and material, two asymptotic cases and an intermediate case can be identified for a given wavelength:
\begin{itemize}[leftmargin=2.0em]
 \item[(s)] grains that are geometrically small,
 \item[(t)] grains that are large but optically thin and
 \item[(o)] grains that are optically thick and geometrically large,
\end{itemize}
corresponding to cases 1, 2, 3 in \citet{kataoka+2014}, respectively.
Grains in categories (s) and (t) absorb with all their transparent volume, thus having $C\sbs{abs} \propto s^3$ and $Q\sbs{abs} \propto s$. The opaque grains in category (o) absorb all radiation that is not reflected off of their surface. They absorb (and emit) only in their surface layers, corresponding to $C\sbs{abs} \propto s^2$ and $Q\sbs{abs} \sim 1$.
\citet{bohren+huffman1983} and \citet{kataoka+2014} give the following explicit absorption efficiencies for the three regimes:
\begin{eqnarray}
  Q\sbs{abs,s} &\approx& 4x \text{Im} \left(\frac{m^2 - 1}{m^2 + 2}\right) = \frac{24 nkx}{(n^2 - k^2 + 2)^2 + (2nk)^2} \propto s,\label{eq:Qabs,s}\\
  Q\sbs{abs,t} &\approx& \frac{8kx}{3n} \left(n^3 - (n^2 - 1)^{3/2}\right) \propto s,\label{eq:Qabs,t}\\
  Q\sbs{abs,o} &\sim& 1,\label{eq:Qabs,o}
\end{eqnarray}
where $m \equiv n + \text{i} k$ is the complex refractive index and $x \equiv 2\pi s/\lambda$ the size parameter.
The transition from (s) to (t) occurs around grain sizes where $2ns \approx \lambda$, corresponding to a characteristic size,
\begin{equation}
  s\sbs{c}' = \frac{\lambda}{2n}.\label{eq:scp}
\end{equation}
Regime (o) is reached when the grains become opaque, that is, around $8kx \sim 3n$, corresponding to
\begin{equation}
  s\sbs{c} = \text{max}\left(\frac{3n\lambda}{16\pi k}, s\sbs{c}'\right).\label{eq:sc}
\end{equation}
Two extremes are plausible:
\begin{itemize}[leftmargin=2.0em]
 \item[(a)] a wide transparent regime (t), that is, $s\sbs{c} \gg s\sbs{c}'$
%  \item[(b)] a negligible difference between small and large transparent grains, i.\,e. $y \approx 1$; and
 \item[(b)] a vanishing transparent regime (t) if $Q\sbs{abs}(s\sbs{c}')$ is already close to unity because of a high $A$, in which case $s\sbs{c} \approx s\sbs{c}'$
\end{itemize}
Figure~\ref{fig:s_c} illustrates cases (a) and (b) for $Q\sbs{abs}$ as a function of grain size. Grains becoming opaque while still being small would imply $Q\sbs{abs,s} \gtrsim 1$ and hence $2n/\lambda \lesssim 4A/3$ or $n \lesssim 8\pi k/3$, which is only met for very strongly absorbing materials.

To simplify the expressions for $Q\sbs{abs,s}$ and $Q\sbs{abs,t}$, the dependence on wavelength can be subsumed in the volume absorption coefficient $A$:
\begin{eqnarray}
  Q\sbs{abs,s} &=& \frac{4}{3}sA,\label{eq:Qabs,s2}\\
  Q\sbs{abs,t} &=& \frac{4}{3}sA y,\label{eq:Qabs,t2}
\end{eqnarray}
where $y \equiv Q\sbs{abs,t}/Q\sbs{abs,s}$. A commonly used alternative symbol for $A$ is $\alpha_\lambda$. The characteristic size for the transition from (t) to (o) then becomes
\begin{equation}
  s\sbs{c} = \text{max}\left(\frac{3Q\sbs{abs,o}}{4A y}, s\sbs{c}'\right).\label{eq:sc-simple}
\end{equation}
As $y$ or $A$ increase, $s\sbs{c}$ approaches $s\sbs{c}'$ and the transparent regime vanishes.

For low absorption, $k \ll n$, the ratio between the two regimes is given by
\begin{equation}
  y \approx \frac{(n^2 + 2)^2}{9n^2} \left(n^3 - (n^2 - 1)^{3/2}\right),
%   y \approx \frac{n^3 - (n^2 - 1)^{3/2}}{9n^2(n^2 + 2)^2}.
\end{equation}
which is then independent from $k$.
Over the range $1 < n < 4$, $y$ can be very roughly approximated with $0.75n^2$.
For materials with low refractive index $n \approx 1$, the difference between (s) and (t) vanishes and $A$ becomes identical with the linear attenuation coefficient, which is related to the imaginary part of the refractive index \citep{bohren+huffman1983}:
\begin{equation}
  A \approx \frac{4\pi k}{\lambda},\label{eq:alpha-k}
\end{equation}
which is independent from $n$.
% In this regime, $A \propto \lambda^{-\beta\sbs{s}}$ corresponds to $k \propto \lambda^{1 - \beta\sbs{s}}$.

Assuming a power-law for the grain size distribution,
\begin{equation}
  N(s) = N_0 \left(\frac{s}{s_0}\right)^{-p},
\end{equation}
and inserting equations~(\ref{eq:Qabs,s2}), (\ref{eq:Qabs,t2}) and (\ref{eq:Qabs,o}) into (\ref{eq:Cabstot}) and (\ref{eq:Mtot}),
the dust mass and total absorption cross-section are
\begin{equation}
  M\sbs{d} = \frac{4\pi \rho N_0 s_0^4}{3(4 - p)} \left[\left(\frac{s\sbs{max}}{s_0}\right)^{4 - p} - \left(\frac{s\sbs{min}}{s_0}\right)^{4 - p}\right]
\end{equation}
and
\begin{eqnarray}
  C\sbs{abs,total}(\lambda) &=& \frac{4\pi A N_0 s_0^4}{3(4 - p)} \left[\left(\frac{s\sbs{c'}}{s_0}\right)^{4 - p} - \left(\frac{s\sbs{min}}{s_0}\right)^{4 - p}\right.\nonumber\\
                                                                                                                &+& \left.y\left(\frac{s\sbs{c}}{s_0}\right)^{4 - p} - y\left(\frac{s\sbs{c'}}{s_0}\right)^{4 - p}\right]\nonumber\\
                            &+& \frac{\pi N_0 s_0^3}{(3 - p)}Q\sbs{abs,o} \left[\left(\frac{s\sbs{max}}{s_0}\right)^{3 - p} - \left(\frac{s\sbs{c}}{s_0}\right)^{3 - p}\right],
\end{eqnarray}
for $s\sbs{min} < s\sbs{c}' < s\sbs{c} < s\sbs{max}$.
The opacity is then
\begin{eqnarray}
  \kappa(\lambda) &=&
  \frac{A}{\rho}
  \left\{
    (1 - y)\left(\frac{s\sbs{c}'}{s\sbs{c}}\right)^{4 - p} - \left(\frac{s\sbs{min}}{s\sbs{c}}\right)^{4 - p} + y\right.\nonumber\\
    &-& \left.\frac{4 - p}{p - 3}\frac{3Q\sbs{abs,o}}{4A s\sbs{c}}\left[\left(\frac{s\sbs{max}}{s\sbs{c}}\right)^{3 - p} - 1\right]
  \right\}\nonumber\\
  &\times&
  \left[
    \left(\frac{s\sbs{max}}{s\sbs{c}}\right)^{4 - p} - \left(\frac{s\sbs{min}}{s\sbs{c}}\right)^{4 - p}
        \right]^{-1},\label{eq:kappa-full}
\end{eqnarray}
where $s_0 = s\sbs{c}$ is assumed.
For a top-heavy size distribution, where big grains dominate both total mass and total cross-section,
\begin{equation}
  \kappa(\lambda) \stackrel{p \ll 3}{\longrightarrow} \frac{3Q\sbs{abs,o}}{4\rho s\sbs{max}} \frac{4 - p}{3 - p},\label{eq:kappa-low-p}
\end{equation}
which inherits only the dependence of $Q\sbs{abs,o}$ on wavelength, roughly leading to $\beta \rightarrow 0$ for $p \ll 3$.
For a bottom-heavy size distribution, where small grains dominate both total mass and total cross-section,
\begin{equation}
  \kappa(\lambda) \stackrel{p \gg 4}{\longrightarrow} \frac{A}{\rho},\label{eq:kappa-high-p}
\end{equation}
which is independent of $p$. A power-law approximation to the volume absorption coefficient,
\begin{equation}
  A = A_0 \left(\frac{\lambda}{\lambda_0}\right)^{-\beta\sbs{s}},\label{eq:alpha}
\end{equation}
then leads to $\beta \rightarrow \beta\sbs{s}$ for $p \gg 4$.

In the typical intermediate regime, $3 < p < 4$, where small grains dominate total cross-section and big grains the total mass, $\kappa$ can be simplified to
\begin{equation}
  \kappa(\lambda) \approx
  \frac{A}{\rho}
  \left\{
    y + (1 - y)\left(\frac{s\sbs{c}'}{s\sbs{c}}\right)^{4 - p} + \frac{4 - p}{p - 3}\frac{3Q\sbs{abs,o}}{4A s\sbs{c}}
  \right\}
  \left(\frac{s\sbs{max}}{s\sbs{c}}\right)^{p - 4}.\label{eq:kappa-intermediate}
\end{equation}

With Eq.~(\ref{eq:sc-simple}), the two cases (a) and (b) imply the opacities
\begin{eqnarray}
  \kappa\sbs{a}(\lambda) &\approx&
  \frac{yA}{\rho}
  \left\{
    1 + \frac{4 - p}{p - 3}
  \right\}
  \left(\frac{s\sbs{max}}{s\sbs{c}}\right)^{p - 4}\nonumber\\
  &\propto& Q\sbs{abs,o} \times s\sbs{c}^{3 - p}~\propto~Q\sbs{abs,o} \times \lambda^{-\beta\sbs{s}(p - 3)}
        \label{eq:kappa-a}
\end{eqnarray}
for a wide transparent regime and
\begin{eqnarray}
  \kappa\sbs{b}(\lambda) &\approx&
  \frac{A}{\rho}
  \left\{
    1 + \frac{4 - p}{p - 3}\frac{3Q\sbs{abs,o}}{4A s\sbs{c}'}
  \right\}
  \left(\frac{s\sbs{max}}{s\sbs{c}'}\right)^{p - 4}\nonumber\\
        &\propto& Q\sbs{abs,o} \times (s\sbs{c}')^{3 - p}~\propto~Q\sbs{abs,o} \times \lambda^{3 - p}
        \label{eq:kappa-b}
\end{eqnarray}
for a direct transition from small to opaque, respectively.
For $y \approx 1$ and $Q\sbs{abs,o} \approx \text{const}$, Eq.~(\ref{eq:kappa-a}) is equivalent to Eq.~(8) in \citet{draine2006}.

The combination of equations~(\ref{eq:kappa-a}) and (\ref{eq:gamma-beta}) results in Eq.~(\ref{eq:draine}), a relation between observed spectral index $\alpha\sbs{mm}$ and size distribution index $p$ that depends on the material through $\beta\sbs{s}$. For case~(b), the combination of Eqs.~(\ref{eq:kappa-b}) and (\ref{eq:gamma-beta}) results in
\begin{equation}
  p = 3 + \alpha\sbs{mm} - \alpha\sbs{Pl},\label{eq:alternative}
\end{equation}
which depends only weakly on the material through $\alpha\sbs{Pl}$ and the grain temperature.

\begin{table}
\centering
\caption{\label{tab:variables}Alphabetic list of recurring symbols.}
\begin{tabular}{rl}
\hline
    Symbol & Meaning \\
\hline
    (a), a & wide transparent regime\\
    $A$ & volume absorption coefficient\\
    $A_0$ & normalization constant for $A(\lambda)$\\
    $\alpha\sbs{mm}$ & measured spectral index at mm wavelengths ($F_\nu$)\\
    $\alpha\sbs{Pl}$ & spectral index of underlying Planck function ($B_\nu$)\\
    (b), b & no transparent regime\\
    $\beta$ & spectral index of $\kappa$\\
    $\beta\sbs{eff}$ & coefficient of linear fit to $\alpha\sbs{mm}(p)$\\
    $\beta\sbs{s}$ & spectral index of $A$\\
    $B_\nu$ & Planck function\\
    $C\sbs{abs}$ & absorption cross-section per grain\\
    $C\sbs{abs,total}$ & total, disc-integrated absorption cross-section\\
    $d$ & distance to observer\\
    $F_\nu$ & spectral flux density (per frequency interval)\\
    $k$ & imaginary part of complex refractive index\\
    $\kappa$ & mass absorption coefficient (or opacity)\\
    $\lambda$ & wavelength\\
    $\lambda_0$ & reference wavelength for $A(\lambda)$\\
    $M$ & grain mass\\
    $M\sbs{d}$ & total disc mass\\
    $n$ & real part of complex refractive index\\
    $N$ & grain number per size interval\\
    $N_0$ & normalization constant for $N(s)$\\
    $\nu$ & frequency\\
    (o), o & opaque (grains)\\
    $p$ & size distribution index\\
    $\Delta p$ & vertical intercept of linear fit to $p(\alpha\sbs{mm}) - 3$\\
    $Q\sbs{abs}$ & absorption (and emission) efficiency\\
    $r$ & distance from star to grain\\
    $\rho$ & bulk grain density\\
    (s), s & small (grains)\\
    $s$ & grain radius\\
    $s_0$ & reference grain radius for $N(s)$\\
    $s\sbs{c}'$ & radius for transition from small to transparent\\
    $s\sbs{c}$ & radius for transition from transparent to opaque\\
    (t), t & transparent (grains)\\
    $T$ & grain temperature\\
    $x$ & grain size parameter\\
    $y$ & ratio between $Q\sbs{abs}$ for (t) and (s) grains\\
    \hline
\end{tabular}
\end{table}

\section{Comparison with numerical results}\label{sec:Numerics}
In this section, the assumptions made for the basic analytic approximations are tested against numerical models for different materials. Absorption and emission are calculated assuming homogeneous, compact spheres. For grains where $2\pi s n < 10^4 \lambda$, Mie theory as implemented by \citet{wolf+voshchinnikov2004} is used. For larger grains the geometrical optics algorithm by \citet{bohren+huffman1983} is used.

\subsection{Materials and range of parameters}
\begin{figure}
  \centering
  \includegraphics{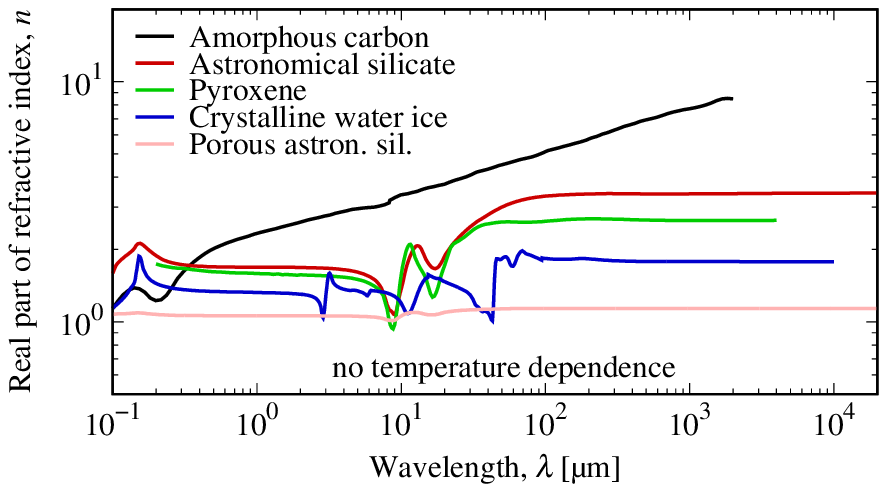}

  \includegraphics{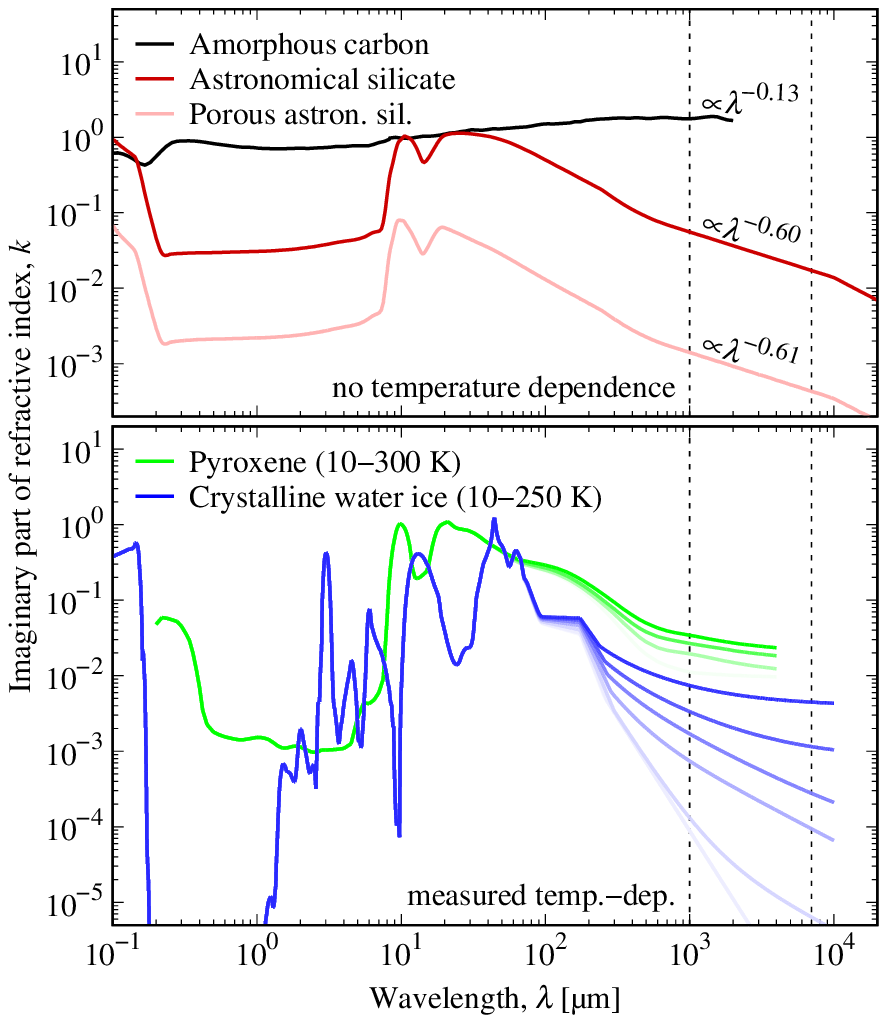}
  \caption{Refractive indexes of the five materials discussed: (top) real and (middle, bottom) imaginary parts. The legends are ordered vertically according to the order at long wavelengths. The vertical dashed lines delimit the wavelength range over which the labelled slopes are fitted.\label{fig:k}}
\end{figure}

A set of five different materials is assumed: compact astronomical silicate \citep{draine2006} with assumed bulk density $\rho = 3.3~\text{g/cm}^3$, porous astronomical silicate \citep[mixing rule:][]{bruggeman1935} with a filling factor $f = 10\,\%$ and $\rho = 0.33~\text{g/cm}^3$, amorphous carbon \citep{zubko+1996} with $\rho = 2~\text{g/cm}^3$, pyroxene (H. Mutschke and P. Mohr, private comm., manuscript in preparation) with $\rho = 3.3~\text{g/cm}^3$, and crystalline water ice \citep{reinert+2015} with $\rho = 1.0~\text{g/cm}^3$.
Only the data sets for the latter two materials include a dependence on grain temperature. Figure~\ref{fig:k} shows the real and imaginary parts of the refractive indexes of these materials. 
All materials exhibit more or less pronounced absorption features for wavelengths $\lambda \lesssim 100$~\micron\ and smoother behaviours for $\lambda \gtrsim 100$~\micron. Amorphous carbon is exceptional in that absorption is high across all considered wavelengths, without a significant decline in the (sub)mm range.

All numerical calculations were done for a grid of parameters. Size distribution indexes are varied from 2.6 to 4.2 in steps of 0.1. Maximum grain radii are $s\sbs{max} = 1$~km for carbon and the silicates and $s\sbs{max} = 100$~km for water ice. Minimum grain radii are determined from the radiation pressure blowout limits for main-sequence stars of six different luminosities: 0.0625, 0.25, 1, 4, 16, along with 64 solar luminosities. In between $s\sbs{max}$ and $s\sbs{min}$, the grid of grain radii is spaced logarithmically with factors of $1.25^{1/3}$ between neighbouring points. Distances between dust and star are set to 1.5625, 3.125, 6.25, 12.5, 25, 50, 100, and 200 au.

Two computations were performed for each of the 4080 combinations of size distribution index, luminosity, distance, and material: one with absorption and emission directly defined by the optical properties of the materials and another one were actual materials are only used for determining grain temperatures (for $s > 1$~mm), but emission was then calculated for blackbodies at these temperatures. The latter computations were used to derive $\alpha\sbs{Pl}$ as required in Eqs.~(\ref{eq:draine}) and (\ref{eq:alternative}).

\subsection{Absorption by volume and by geometrical cross-section}
\begin{figure*}
  \centering
  \includegraphics{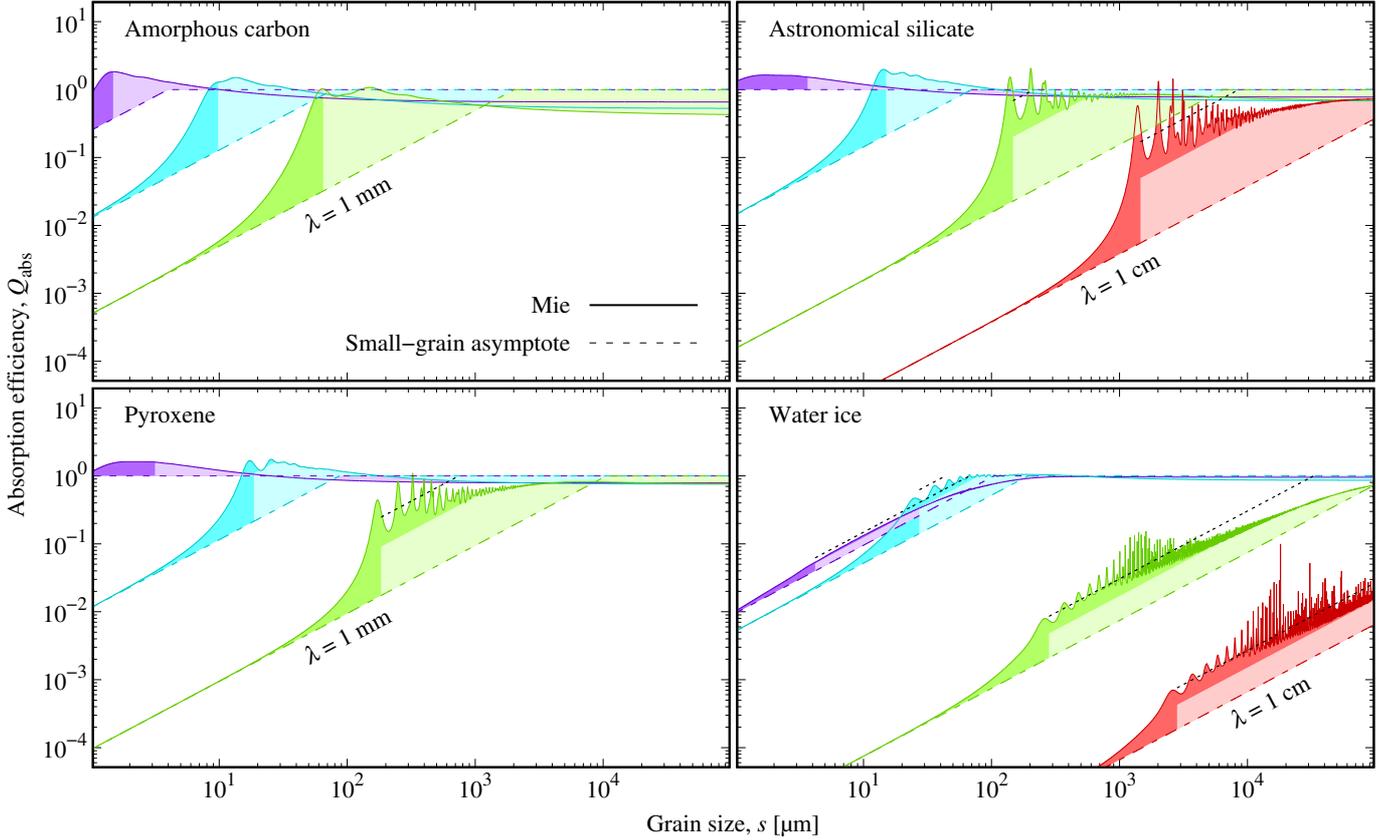}
  \caption{Absorption efficiencies for four materials and a range of wavelengths and grain sizes. Wavelength are separated by factors of 10, the longest wavelengths are labelled. Solid lines indicate Mie results, dashed lines the small-grain asymptote given in Eq.~(\ref{eq:Qabs,s}). The upper boundaries of the light-shaded regions are defined by Eq.~(\ref{eq:Qabs,t}), the left boundaries by Eq.~(\ref{eq:scp}). Black dotted lines follow Eq.~(\ref{eq:Qabs,t}) multiplied by an additional factor $n$ to roughly trace the actual smoothed $Q\sbs{abs}$. Values are given for stellar luminosity $L = L\sbs{Sun}$ and a distance $r = 100$~au.\label{fig:Qabs}}
\end{figure*}

The resulting absorption efficiencies are plotted in Fig.~\ref{fig:Qabs} for a range of wavelengths and grain sizes, illustrating the occurrences of the regimes (s), (t), and (o) as defined in Sect.~\ref{sec:Approximations}. As expected, the opaque regime (o) with $Q\sbs{abs} \sim 1$ is found where $k$ is high and wavelengths $\lambda$ are short compared to grain radii $s$. The transparent regime (t) is found for $s \gtrsim \lambda/(2n)$ but low absorption. As soon as $s \lesssim \lambda/(2n)$, the grains can be considered small, corresponding to regime (s). The transition radius and the ratio $y = Q\sbs{abs,t}/Q\sbs{abs,s}$ depend, notably, on $n$.

Figure~\ref{fig:Qabs} shows that, for silicates and carbon, the transparent regime is missing for $\lambda \lesssim 1$~mm. Absorption transitions directly from small to opaque regime, corresponding to case (b) and Eq.~(\ref{eq:alternative}) from Sect.~\ref{sec:Approximations}. For $\lambda \gtrsim 1$~mm, silicates and carbon show a transparent regime. At these longer wavelengths, a transition to case (a) and Eq.~(\ref{eq:draine}) can be expected.

The assumption of homogeneous, perfectly spherical grains leads to artefacts in the $Q\sbs{abs}$ curves. In the transparent regime, the numerical values for $Q\sbs{abs}$ exceed the values predicted by Eq.~(\ref{eq:Qabs,t}) significantly, showing strong oscillations. This excess amounts on average to an additional factor $\approx n$. This is due to the monochromatic calculations and the idealized shape resulting in internal resonances, which increase absorption and emission. For more irregular grains and integration over a broader range of wavelengths, the curves are expected to become smoother. In the opaque regime, $Q\sbs{abs}$ of amorphous carbon grains converges to $\approx 0.4$, that is, a value significantly below unity. This is caused by the assumed even surfaces and the resulting Fresnel reflection on the surface of this high-$n$ material. Surface structure on sub-wavelength scales could reduce reflection and increase absorption.

\subsection{Opacities}
\begin{figure}
  \centering
  \includegraphics{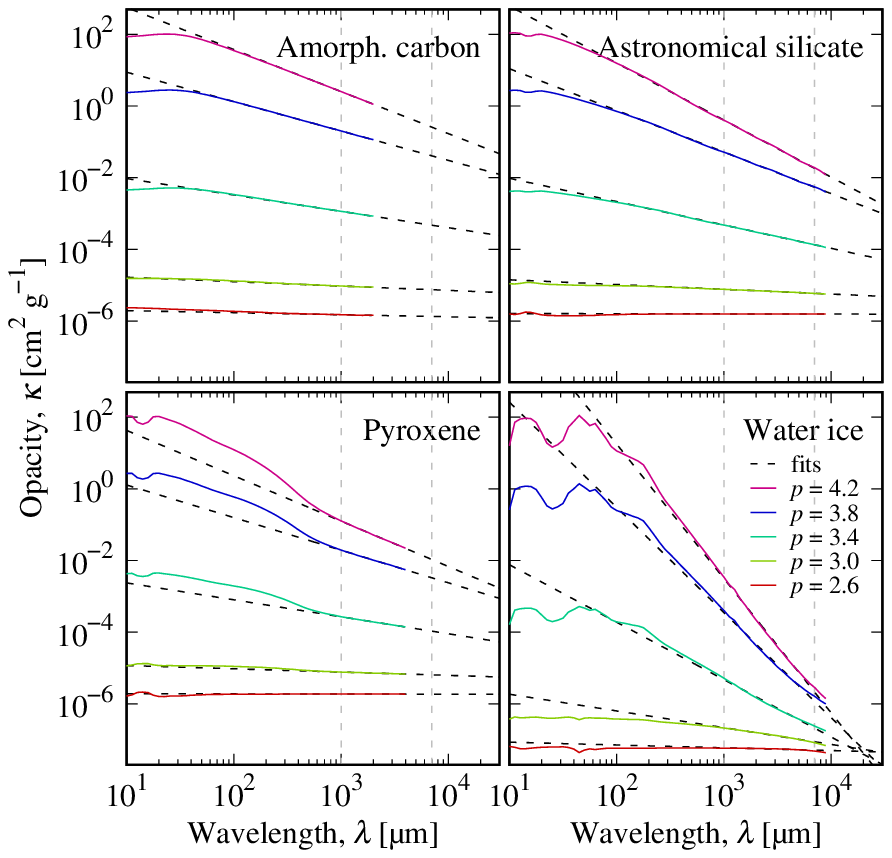}
  \caption{Opacities for four different materials for a range of size distribution slopes and wavelengths. Values are given for stellar luminosity $L = L\sbs{Sun}$, a distance $r = 100$~au, and a total disc mass $M\sbs{d} = 0.1M\sbs{Earth}$. The maximum object radius is set to $s\sbs{max} = 100$~km for water ice and 1~km for carbon and the silicates. Mie results are plotted with solid lines for five size distribution indexes from (top) $p = 4.2$ to (bottom) $p = 2.6$. Sloped dashed lines trace power-law fits to wavelengths from $\lambda_1 = 1$~mm to the shorter of $\lambda_2 = 7$~mm and the maximum wavelength in the tabulated optical data. Vertical dashed lines indicate $\lambda_1$ and $\lambda_2$.\label{fig:kappa-lambda-p}}
\end{figure}

The total mass absorption coefficient or opacity, $\kappa$, is an intermediary between absorption efficiencies and SEDs. Figure~\ref{fig:kappa-lambda-p} shows results for four different materials as functions of size distribution indexes and wavelengths. These curves partly inherit the absorption features seen for $k$, but they are smoother because they tend to be dominated by grains that are rather opaque. The opacities of astronomical silicate and amorphous carbon follow smooth power-laws already for $\lambda \gtrsim 100$~\micron. The data sets for pyroxene and water ice have measured changes in slope for $\lambda \gtrsim 100$~\micron, which lead to a corresponding significant departure of $\kappa$ from a simple power law.

All four materials show near-constant opacities for shallow size distributions around $p = 3$ because in this case, big and opaque grains near to $s\sbs{max}$ not only dominate the total mass but also contribute significantly to the total cross-section. The disc then has the same effective cross-section at all wavelength. The remaining slope for carbon at $p = 2.6$ is caused by high $k$, and $n$ increasing with $\lambda$. As a result, reflectance increases and $Q\sbs{abs,o}$ decreases. For the silicates and water ice, $k$ is lower and $n$ constant.

Towards size distribution slopes $p \geq 4$, smaller grains become more and more important. Given the same total mass $M\sbs{d}$, the total cross-section $C\sbs{abs,total}$ increases, leading to higher values for $\kappa$ overall. However, the small grains' effective cross-section more strongly depends on wavelength, leading to steeper opacity slopes for steeper size distributions, as expected from both Eqs.~(\ref{eq:kappa-a}) and (\ref{eq:kappa-b}). Around wavelengths of a few tens of microns, $k$ is comparably high for all materials, and so is $\kappa$.

The wavelengths at which the $\kappa$ curves for different size distribution slopes converge are the wavelengths at which the biggest assumed grains become transparent, that is, the wavelengths at which $s\sbs{c}(\lambda) = s\sbs{max}$. At yet longer wavelengths, $\kappa$ would become insensitive to $p$. For carbon and the silicates, the extrapolated curves converge at wavelengths of several metres or even kilometres. For water ice at $\sim 10$~K they converge already around $\lambda \sim 1$~dm because of the very low absorption, despite the increased $s\sbs{max}$ for this material.
 
\begin{figure}
  \centering
  \includegraphics{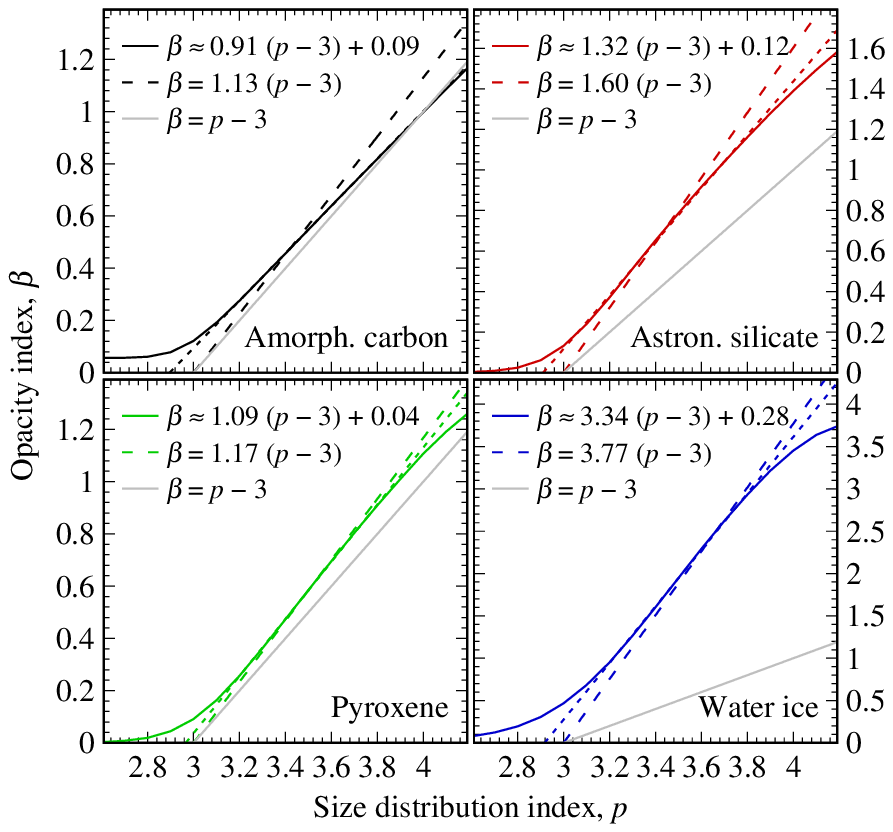}
  \caption{Relation between the power-law indices of the grain size distributions and the total opacities between $\lambda_1 = 1$ mm and $\lambda_2 = 7$ mm for four materials at $r = 100$~au around a star with $L = L\sbs{Sun}$. Solid coloured lines show numerical results. Dotted lines of the same colours show linear fits for $3.2 \leq p \leq 3.8$, with fit coefficients indicated in the figure keys. Dashed lines show the analytic approximation according to Eq.~(\ref{eq:kappa-a}). The solid grey lines trace the approximate Eq.~(\ref{eq:kappa-b}).\label{fig:beta-p}}
\end{figure}

Figure~\ref{fig:beta-p} compares the $\beta(p)$ dependencies expected from the Mie calculations with the analytic approximations in equations~(\ref{eq:kappa-a}) and (\ref{eq:kappa-b}). Neither of the two approximations can reproduce the numerical results over the full range of size distribution slopes for all materials. While the solid curve for amorphous carbon approaches the grey line that follows Eq.~(\ref{eq:kappa-b}), the results for pyroxene, water ice, and astronomical silicate are closer to the dashed lines that represent Eq.~(\ref{eq:kappa-a}). The slopes of the regression lines for the two silicates are only halfway between that of Eqs.~(\ref{eq:kappa-a}) and (\ref{eq:kappa-b}), meaning that the sensitivity of $\beta$ to differences in $p$ is also halfway between cases (a) and (b). These findings for the four materials are consistent with their refractive indexes in the mm wavelength regime. The high $n$ and $k$ of carbon puts it in category (b), the low absorption of pure water ice leads to case (a), with the two silicates remaining in between.

\begin{figure}
  \centering
  \includegraphics{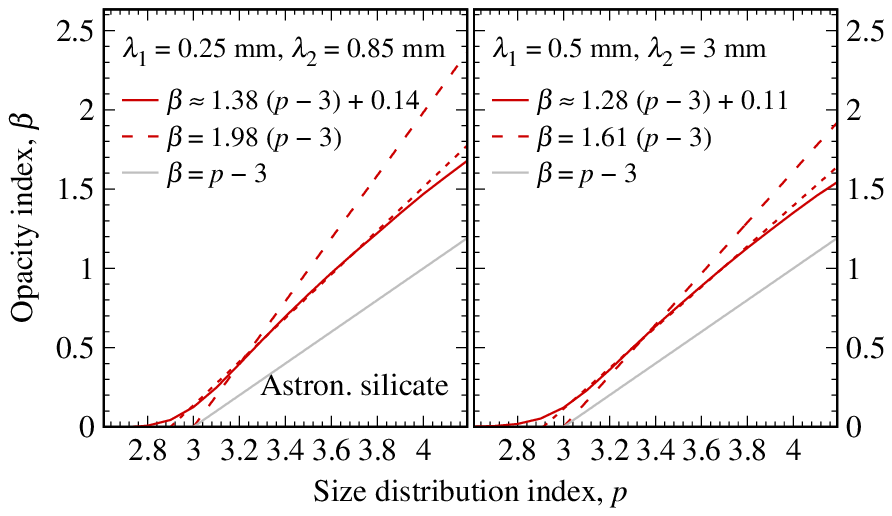}
  \caption{Same as the top-right panel in Fig.~\ref{fig:beta-p} (Astronomical Silicate) but for two combinations of shorter wavelengths, as indicated.\label{fig:beta-p-23}}
\end{figure}

Taking Astronomical Silicate as an example, Fig.~\ref{fig:beta-p-23} shows how the results shift more towards case (b) for shorter, submm wavelengths. In this wavelength range absorption is higher for all materials, resulting in a narrower transparent regime (t).

\begin{figure}
  \centering
  \includegraphics{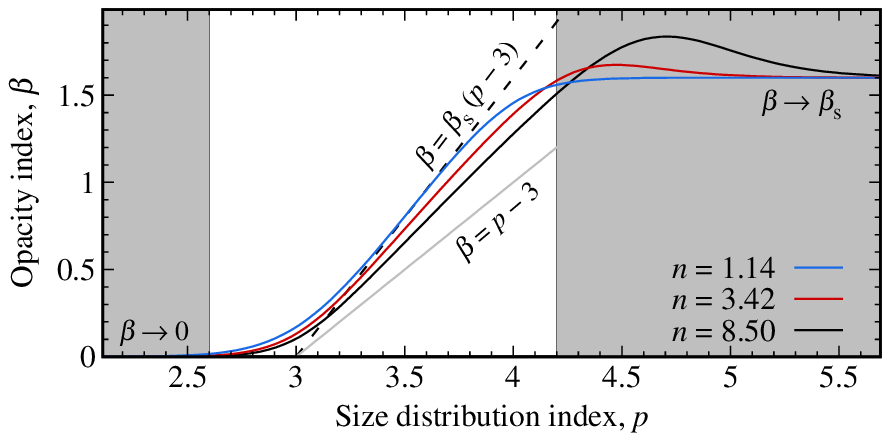}
  \caption{Slope of the opacity as derived from Eq.~(\ref{eq:kappa-full}) for the following parameters: $\lambda = 3$~mm, $\beta\sbs{s} = 1.6$, $A_0 = 2/\text{cm}$, $\lambda_0 = 1$ mm, $\rho = 3.5~\text{g~cm}^{-3}$, $s\sbs{min} = 0.5$~\micron, $s\sbs{max} = 100$~m, $Q\sbs{abs,o} = 1$, $y = 0.75 n^2$, with (black) $n = 8.5$, (red) $n = 3.4$, and (blue) $n = 1.14$. In the range $3 < p < 4$, this order of colours reflects the vertical order of the corresponding lines, bottom to top. The dashed black and solid grey lines follow Eqs.~(\ref{eq:kappa-a}) and (\ref{eq:kappa-b}), respectively.\label{fig:beta-p-analytic}}
\end{figure}

\begin{figure}
  \centering
  \includegraphics{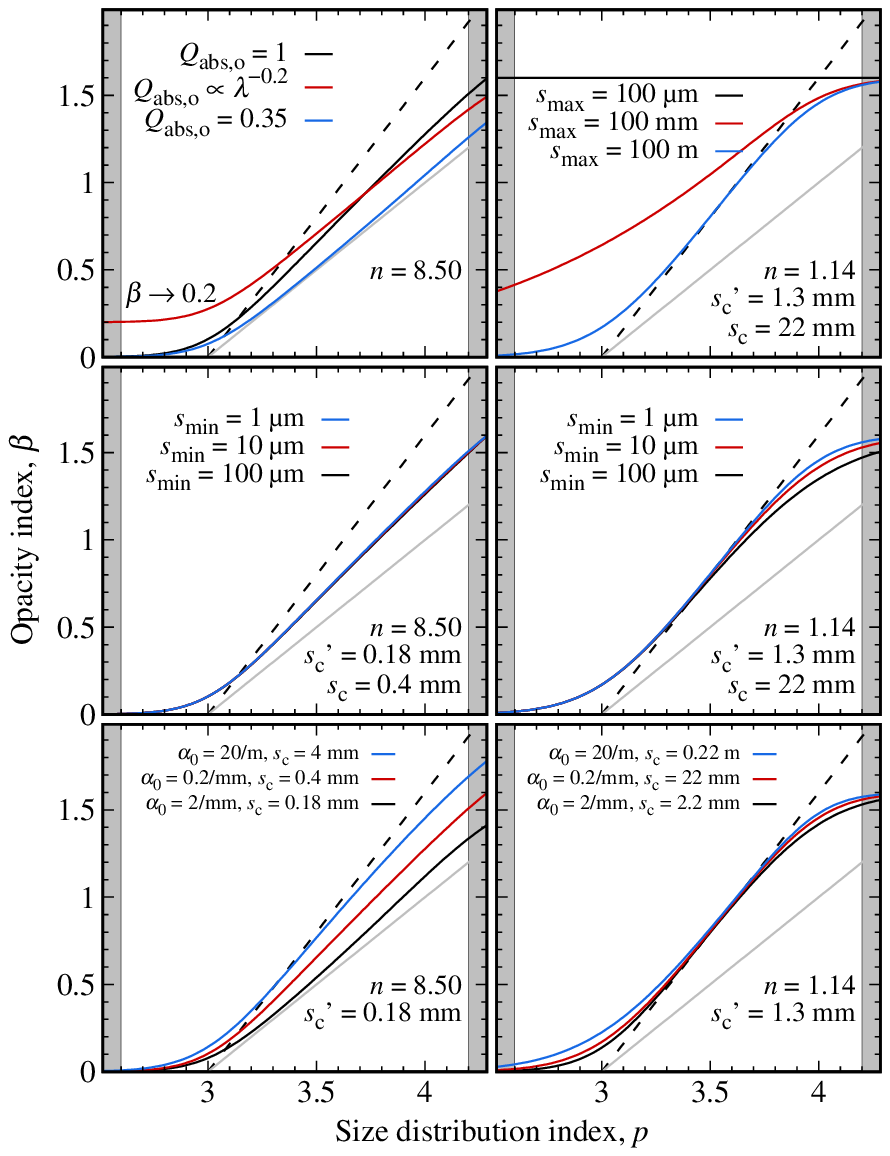}
  \caption{Same as Fig.~\ref{fig:beta-p-analytic}, but with (top left) $Q\sbs{abs,o}$, (top right) $s\sbs{max}$, (middle) $s\sbs{min}$, and (bottom) $A_0$ varied. The left panels show results for high $n$, the right panels for low $n$. The legends are ordered vertically according to the order at $p \approx 4$.\label{fig:beta-p-variations}}
\end{figure}

Figure~\ref{fig:beta-p-analytic} shows that the general shape of the $\beta(p)$ relation is well-reproduced by Eq.~(\ref{eq:kappa-full}). The slopes of the near-linear sections around $p=3.5$ are determined by the refractive index, with shallower slopes for higher $n$. Around $p\approx 3$ and $p \approx 4$ the curves turn towards the asymptotic Eqs.~(\ref{eq:kappa-low-p}) and (\ref{eq:kappa-high-p}), respectively. For materials with high $n$, $\beta$ can exceed $\beta\sbs{s}$ for $p > 4$, while still converging to $\beta\sbs{s}$ for $p \rightarrow \infty$.
Figure~\ref{fig:beta-p-variations} illustrates further dependencies of Eq.~(\ref{eq:kappa-full}) on the material properties and grain size ranges. A lower absorption efficiency $Q\sbs{abs},o$, caused by a more reflective surface due to a greater $n$, reduces the step factor $y$, narrows the transparent regime and brings the results closer to Eq.~(\ref{eq:alternative}). A wavelength dependence $Q\sbs{abs,o}(\lambda) \propto \lambda^{-\xi}\neq \text{const}$ in the mm wavelength range, caused by $n(\lambda) \neq \text{const}$, results in a non-zero $\beta \rightarrow \xi$ for $p < 3$. This effect is also seen in the numerical results for amorphous carbon in the top-left panel of Fig.~\ref{fig:beta-p}. If the maximum grain radius is comparable to or less than the characteristic grain size $s\sbs{c}$, the opaque regime vanishes and $\beta(p)$ becomes shallower. If $s\sbs{max} \leq s\sbs{c}'$, only the small-grain regime remains and $\beta = \beta\sbs{s} = \text{const}$. This is relevant for observed young protoplanetary discs where grain growth may not have reached $s\sbs{max} > s\sbs{c}$ yet \citep[e.\,g.][]{kruegel+siebenmorgen1994,natta+testi2004,draine2006}. It is also relevant for modelled discs, where systematic errors are introduced if the assumed $s\sbs{max}$ is too small.

The greater the minimum grain radius $s\sbs{min}$, the fewer small grains are present and the steeper is the size distribution required for the small grains to dominate. With increasing $s\sbs{min}$, $\beta(p)$ flattens (and $p(b)$ steepens) around $p \approx 4$ because the small-grain regime $\beta\sbs{s}$ is approached only for higher $p$. This dependence of $\beta(p)$ on $s\sbs{min}$ translates to a dependence on stellar luminosity if the latter is high enough to set a lower size cutoff via radiation pressure blowout of the smallest grains.
For materials with high $n$ the effect of different $s\sbs{min}$ is smaller, as long as $s\sbs{min} < s\sbs{c}'$. Because of the jump by a great factor $y$ from $Q\sbs{abs,s}$ to $Q\sbs{abs,t}$, the small grains play a minor role and the exact location of their lower size cut-off is not as important. Only as $s\sbs{min}$ approaches or exceeds $s\sbs{c}'$ or $s\sbs{c}$ does the $\beta$ relation react by converging quickly towards a constant $\beta \approx 0$. This is the case for carbon grains around high-luminosity stars, where radiation pressure blowout can eliminate small and transparent grains completely. For materials with low $n$, small grains and their lower size cutoff are more important. The lower $s\sbs{min}$, the more abundant are the small grains and the quicker is the convergence $\beta(p) \rightarrow \beta\sbs{s}$ around p$\approx 4$.

A lower volume absorption coefficient, as represented by $A_0$ in the bottom row of Fig.~\ref{fig:beta-p-variations}, brings $\beta(p)$ closer to Eq.~(\ref{eq:draine}) in the range $3 < p < 4$ for both high-$n$ and low-$n$ materials. For materials with high $n$, this is due to a lower $A$ causing a greater characteristic grain radius, $s\sbs{c}$ , and a wider transparent regime. This effect does not depend on $s\sbs{min}$ or $s\sbs{max}$. For materials with low $n$, where the transparent regime is less important, a lower $A$ brings $s\sbs{c}$ closer to $s\sbs{max}$ and narrows the opaque regime. This effect for low $n$ is similar to that shown in the top-right panel of Fig.~\ref{fig:beta-p-variations}. It disappears for $s\sbs{max} \gg s\sbs{c}$.


\subsection{Spectral index}\label{sec:recipe}
\begin{figure}
  \centering
  \includegraphics{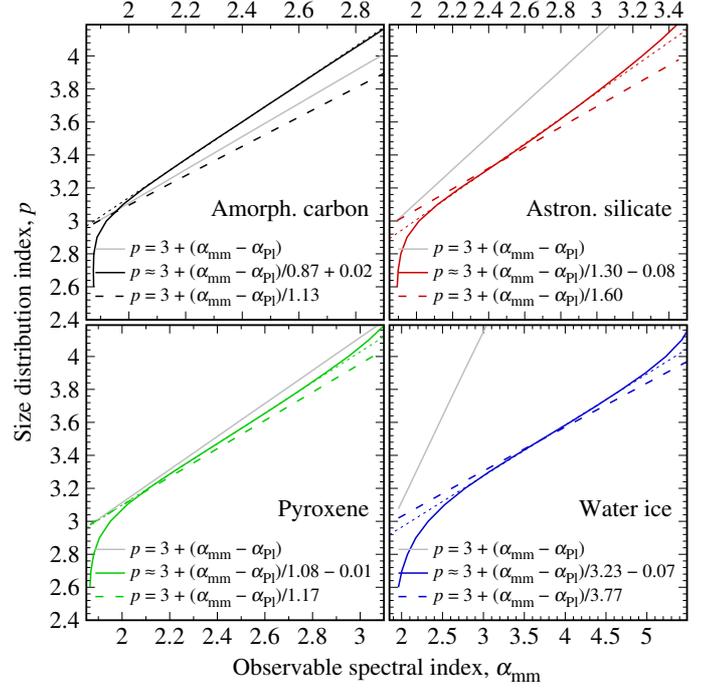}
  \caption{Relation between the power-law indexes of the grain size distribution and the observed spectrum for four materials. Parameters and line meanings are the same as in Fig.~\ref{fig:beta-p}. Ranges for the horizontal axes are different for the four panels, indicated above the top panels and below the bottom panels.\label{fig:alpha-p}}
\end{figure}

For a direct link between observed spectral index and size distribution index, SEDs are computed for the range of parameters. As long as temperatures do not vary strongly across the relevant grain sizes, SEDs and their slopes simply reflect the opacity index and the slope of the blackbody function at that temperature. In Fig.~\ref{fig:alpha-p} the analytic predictions for size distribution index $p$ as a function of the observable spectral index in the mm regime, $\alpha\sbs{mm}$, are compared to the numerical results. As expected from the close relation between $\beta$ and $\alpha\sbs{mm}$, the findings for the $p(\alpha\sbs{mm})$ relation reflect those for $\beta(p)$ and the curves shown in Fig.~\ref{fig:beta-p}. The full numerical solution typically falls in between cases (a) and (b), represented by Eqs.~(\ref{eq:alternative}) and (\ref{eq:draine}), respectively. With $\alpha\sbs{mm}$ (and $\beta$) being less sensitive to $p$ than predicted for case (b), $p$ is in turn more sensitive to differences or uncertainties in $\alpha\sbs{mm}$. All else being equal, the size distribution indexes for systems with different $\alpha\sbs{mm}$ differ more strongly than expected from Eq.~(\ref{eq:draine}). 

\begin{figure}
  \centering
  \includegraphics{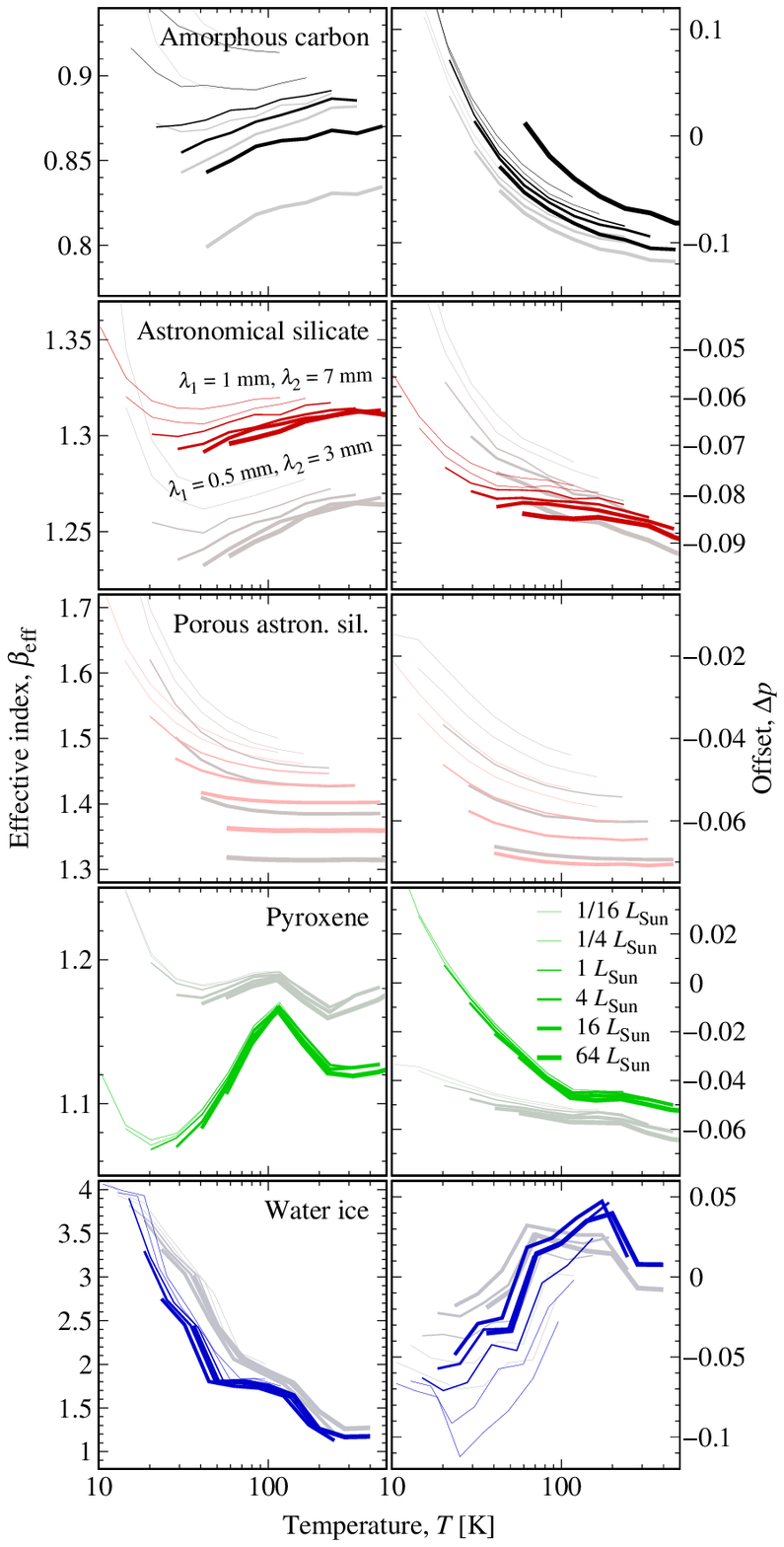}
  \caption{Regression coefficients for linear fits to the modelled relation between observable spectral slope $\alpha\sbs{mm}$ and the size distribution index $p$ as defined in Eq.~(\ref{eq:alpha-p}). Coloured lines indicate values for longer wavelengths, grey lines for short wavelengths. Line width increases with stellar luminosity from $L = 0.0625 L\sbs{Sun}$ to $L = 64 L\sbs{Sun}$.\label{fig:all-alpha-p}}
\end{figure}

The curves for $p(\alpha\sbs{mm})$ are roughly linear for $3.2 \leq p \leq 3.8$ and can be approximated well by
\begin{equation}
  p \approx 3 + \frac{\alpha\sbs{mm} - \alpha\sbs{Pl}}{\beta\sbs{eff}} + \Delta p\label{eq:alpha-p}
\end{equation}
in that range, underlining the similarity with Eqs.~(\ref{eq:alternative}) and (\ref{eq:draine}). The coefficients $\beta\sbs{eff}$ and $\Delta p$ for all assumed combinations of materials, distances, and stellar luminosities can be read from Fig.~\ref{fig:all-alpha-p}. The dependence on stellar luminosity is mostly weak because grains near to $s\sbs{min}$, as defined by radiation pressure blowout, contribute only slightly to mm emission. One exception is carbon with its high absorption efficiency, which shifts the characteristic grain size close to or below the lower size limit. At $L = 64~L\sbs{Sun}$, the values for $\beta\sbs{eff}$ are around 0.4, outside the plotted range. The other exception is the porous silicate, where $n$ and $\rho$ are low, the blowout limit higher, the small-grain regime more important, and the dependence on stellar luminosity more pronounced.

The following recipe can then be used for determining the size distribution indexes from observed spectral energy distributions: 1) pick a material (mix) that is appropriate for your system; 2) derive the temperature of the big grains; 3) derive $\alpha\sbs{Pl}$ for that temperature and your pair of wavelengths; 4) read $\beta\sbs{eff}$ and $\Delta p$ off of Fig.~\ref{fig:all-alpha-p}; and 5) calculate $p$ from Eq.~(\ref{eq:alpha-p}).

Figure~\ref{fig:all-alpha-p} also shows a notable increase of $\beta\sbs{eff}$ for temperatures below $\approx 30$ K for all materials. This dependence on temperature is where the behaviours of $\alpha\sbs{mm}(p)$ and $\beta(p)$ differ. The following section has a more detailed discussion.

\subsection{Grain temperature}\label{sec:temp}
\begin{figure}
  \centering
  \includegraphics{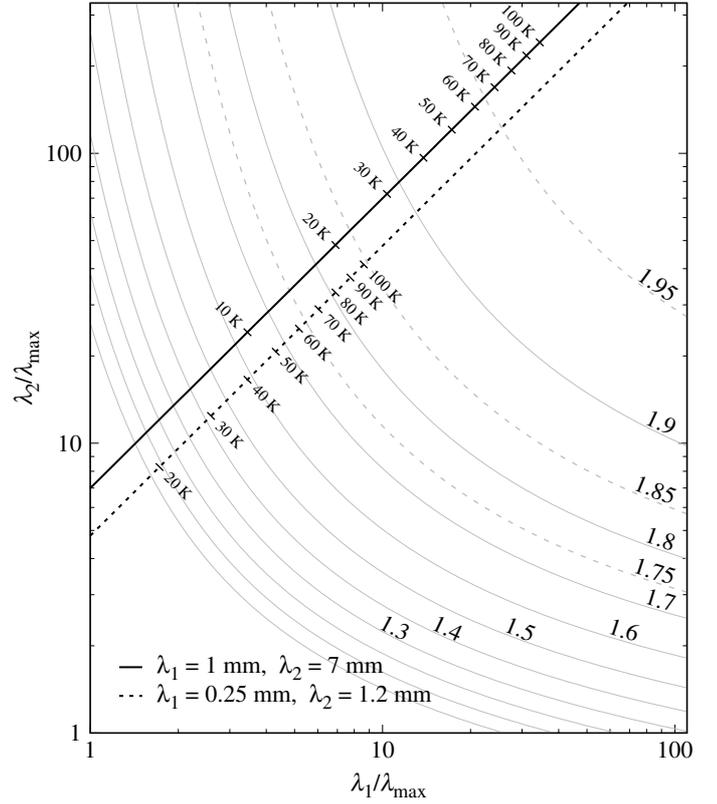}
  \caption{Blackbody spectral indices as a function of combinations of wavelengths, normalized to temperature-dependent blackbody peak wavelengths $\lambda\sbs{max}$.\label{fig:alpha-pl}}
\end{figure}

Grain temperature influences the relation between size distribution and SED in two ways. First, the optical properties can vary with temperature. The relation between opacity index and size distribution index thus depends on stellar luminosity and disc radius. This variability in $\beta\sbs{eff}$ is visible in Fig.~\ref{fig:all-alpha-p}. The temperature-dependent water ice shows stronger variability in $\beta\sbs{s}$, $\beta\sbs{eff}$, and $\Delta p$. The slope $\beta\sbs{s}$ of pure water ice at low temperatures is so steep that even a very steep SED could correspond to a moderate $p$. For carbon, the two variants of Astronomical Silicate, and pyroxene, $\beta\sbs{s}$ does not or only weakly depend on temperature, and $\beta\sbs{eff}$ is less variable. Second, the Planck function in Eq.~(\ref{eq:basic}) and the corresponding power-law index $\alpha\sbs{Pl}$ depend on temperature. Figure~\ref{fig:alpha-pl} shows how $\alpha\sbs{Pl}$ varies with observing wavelengths and temperature. The colder the grains and the shorter the wavelengths, the stronger is the deviation of $\alpha\sbs{Pl}$ from the value of 2 found in the Rayleigh limit. Any uncertainty in the physical grain temperature therefore translates to an uncertainty in the derived size distribution index. 

\begin{figure}
  \centering
  \includegraphics{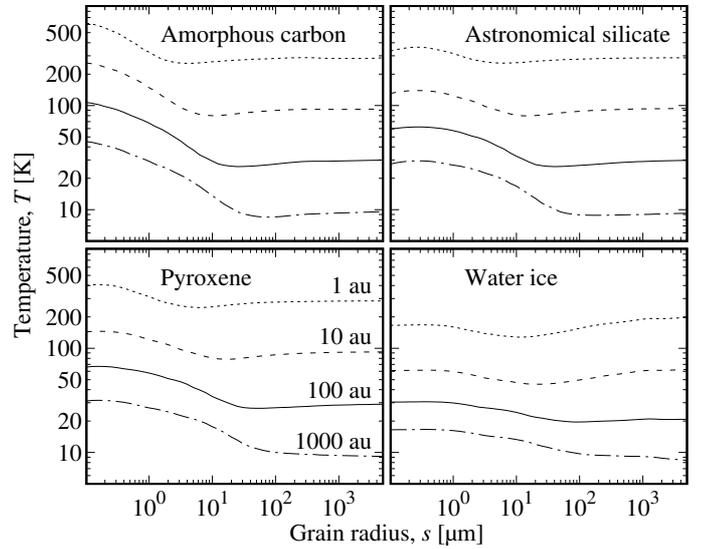}
  \caption{Equilibrium grain temperature as a function of grain size for four materials and four distances around a star of solar luminosity.\label{fig:temps}}
\end{figure}

Figure~\ref{fig:temps} shows the equilibrium temperatures resulting from Eq.~(\ref{eq:temps}). For grains with radii $s \gtrsim 100$~\micron\ temperature is nearly constant. These grains are large enough to emit efficiently because their $Q\sbs{abs}$ is close to unity in the far infrared and submm. Their temperature is no longer a function of size. With the exception of very transparent materials, such as pure water ice, this temperature is close to the blackbody value \citep[see, e.\,g.][]{pawellek+2014}. However, the temperature, which is assumed constant from Eq.~\ref{eq:Fnu} onwards, actually depends on grain size. Smaller grains, which cannot re-emit as efficiently, can be significantly hotter. The top-left panel of Fig.~\ref{fig:temps} shows an extreme case, where the smallest carbon grains at 1000~au are hotter by up to a factor of five. When the size distribution is steep enough, the contribution from small ($\lesssim 100$~\micron) grains to the mm emission is significant. The effective blackbody spectral index, determined from the weighted average over all grain sizes, increases for higher $p$ because of the increased abundance of smaller, hotter grains. The $\alpha\sbs{mm}(p)$ relation inherits that dependence and steepens, while $p(\alpha\sbs{mm})$ flattens, corresponding to a higher $\beta\sbs{eff}$. This effects becomes more pronounced for distant discs or less luminous stars, where temperature and spectral index are lower in general and the relative spread in $T$ is higher. Hence, the $\beta\sbs{eff}$ curves in Fig.~\ref{fig:all-alpha-p} rise as temperatures drop below $\sim 30$~K.

For spatially resolved circumstellar rings, the temperature of big grains can be computed from the distance to the star and the stellar luminosity. In the more common case, where the disc is only barely resolved or not at all, distance and temperature can only be assessed through the SED. However, this procedure is partly degenerate because the SED is already a convolute of size and temperature distributions. Determining grain temperature independently from the size distribution can be difficult.
Near to the flux density peak, which is around $\lambda \lesssim 100$~\micron\ for cold debris discs, grains with $s < 100$~\micron\ or even $s \ll 100$~\micron\ contribute or dominate as long as $p \gtrsim 3$. These warmer grains shift the emission peak to shorter wavelengths, masking the blackbody curve of the bigger grains that dominate at longer wavelengths. The overestimated temperatures result in overestimated $\alpha\sbs{Pl}$ and underestimated $p$ for given $\alpha\sbs{mm}$. As shown in Fig.~\ref{fig:alpha-pl} for observations at 1~mm and 7~mm, assumed temperatures of 20~K or 50~K instead of 10~K would result in $\alpha\sbs{Pl} = 1.83$ or $\alpha\sbs{Pl} = 1.93$ instead of $\alpha\sbs{Pl} = 1.64$. For $\beta\sbs{eff} \approx 1.3$ and a given $\alpha\sbs{mm}$, this would correspond to errors in $p$ of $-0.15$ or $-0.22$, respectively. The SED analysis can be further complicated by the different temperatures involved for debris or protoplanetary discs that are very extended or composed of two or more separate rings of comparable luminosity.

\subsection{Grain porosity}\label{sec:porosity}
\begin{figure}
  \centering
  \includegraphics{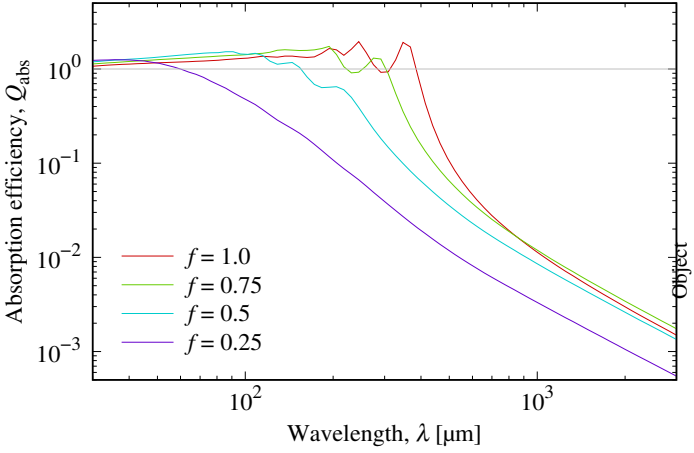}
  \caption{Absorption efficiencies for porous spheres of Astronomical Silicate with radii $s = 50$~\micron\ and filling factors $f = 1$, 0.75, 0.5, 0.25 (top to bottom at $\lambda \approx 400$~\micron).\label{fig:dda-porosity}}
\end{figure}

To illustrate the general impact of porosity, \texttt{ddscat} in version 7.3.2 \citep{draine+flatau1994,draine+flatau2013x} is used to model spherical grains of Astronomical Silicate with filling factors $f = 1$, 0.75, 0.5, and 0.25, corresponding to porosities of 0, 25, 50, and 75\,\%. The grain radius is fixed at $s = 50$~\micron. The size of the individual dipoles was is to 0.5~\micron, much smaller than the wavelengths in the considered range $30~\micron < \lambda < 3~\text{mm}$. The resulting $Q\sbs{abs}(\lambda)$ curves in Fig.~\ref{fig:dda-porosity} show a trend towards lower $Q\sbs{abs}$ values for lower filling factors over a wide range of wavelengths, consistent with lower absorption for less dense material. At long wavelengths, the combined dependence of $Q\sbs{abs}$ on $n$ and $k$ given in Eq.~\ref{eq:Qabs,s} is reflected. At shorter wavelengths, in the opaque regime, $Q\sbs{abs}$ is more affected by $n$. There, lower filling factors result in lower $n$, reduced reflection, and hence, increased absorption.

The transition from (s) small and transparent to (o) large and opaque is also affected by porosity. For $f = 1$, $n$ is high enough for a direct transition from regime (o) to (s). For lower $f$ and lower $n$ the intermediate regime (t) becomes notable. With decreasing $f$, the transition from (s) to (t) occurs at decreasing wavelengths because $\lambda \propto n$ for constant $s\sbs{c}'$ (see Eq.~\ref{eq:scp}). The transition from (t) to (o) occurs at $\lambda \propto k/n$, i.\,e. $\lambda$ decreases with decreasing $k$, despite $n$ levelling off at unity.

For $f \rightarrow 0$, as $n \rightarrow 1$, regime (t) becomes indistinguishable from regime (s) because refraction becomes unimportant and the off-setting factor $y$ approaches unity. The $\beta(p)$ relation for porous materials is therefore expected to follow Eq.~(\ref{eq:kappa-a}) more closely compared to their compact counterparts, in line with the steeper curves for lower $n$ seen in Fig.~\ref{fig:beta-p-analytic}. As a result, porous grains exhibit higher values of $\beta\sbs{eff}$ and $\alpha\sbs{mm}$ when compared to compact grains with the same size distribution index $p$. This effect has been shown by \citet{brunngraeber+2017} and is visible in Fig.~\ref{fig:all-alpha-p}. At the same time, the lower bulk density of porous material increases the blowout size, and hence, strengthens the dependence on disc temperature and stellar luminosity. The coefficient $\beta\sbs{eff}$ does therefore not converge towards $\beta\sbs{s}$, but keeps increasing for increasing porosity \citep[cf.][]{brunngraeber+2017}.

\subsection{Application to observed debris discs}\label{sec:Application}

\begin{table*}
\centering
\caption{\label{tab:alpha-data}Observed spectral indexes for a set of debris discs.}
\begin{tabular}{r@{~}lrlllr@{~}c@{~}llrlrr}
\hline
    System & & $L_*$ & Ref. & $\lambda_1$ & $\lambda_2$ & $\alpha\sbs{mm}$ & & & Ref. & $r$ & Ref. & $T$ & $\alpha\sbs{Pl}$ \\
           & & [$L_\odot$] &  & [mm] & [mm] & & & & & [au] & & [K] & \\
\hline
    AU Mic &  & 0.1 & 1 & 1.3 & 9.0 & $<2.46$ & & & 1 & 40 & 11 & 26 & 1.87 \\
    $\epsilon$ Eri &  & 0.3 & 1 & 1.3 & 7.0 & $>2.39$ & & & 7 & 69 & 12 & 26 & 1.87 \\
    HD 61005 &  & 0.5 & 1 & 1.3 & 9.0 & $2.49$ & $\pm$ & 0.08 & 1 & 67 & 13 & 30 & 1.89 \\
    HD 107146 & i & 1.0 & 1 & 1.25 & 7.0 & $2.55$ & $\pm$ & 0.11 & 8 & 50 & 14 & 41 & 1.92\\
    HD 107146 & o & 1.0 & 1 & 1.25 & 7.0 & $2.55$ & $\pm$ & 0.11 & 8 & 116 & 14 & 27 & 1.88 \\
    HD 377 & & 1.0 & 1 & 0.87 & 9.0 & $>2.39$ & & & 1 & 50 & 15 & 42 & 1.92 \\
    HD 105 & & 1.2 & 2 & 0.87 & 9.0 & $2.41$ & $\pm$ & 0.16 & 9 & 85 & 2 & 33 & 1.90 \\
    q$^1$ Eri & & 1.2 & 1 & 0.87 & 7.0 & $2.94$ & $\pm$ & 0.10 & 1 & 85 & 15 & 33 & 1.90 \\
    HD 104860 & & 1.4 & 1 & 1.3 & 9.0 & $3.08$ & $\pm$ & 0.23 & 1 & 110 & 16 & 31 & 1.89 \\
    AK Sco & & 3.0 & 1 & 1.3 & 7.0 & $2.62$ & $\pm$ & 0.03 & 1 & 14 & 1 & 103 & 1.97 \\
    HD 15115 & i & 3.3 & 1 & 1.3 & 9.0 & $2.75$ & $\pm$ & 0.15 & 1 & 48 & 17 & 57 & 1.92 \\
    HD 15115 & o & 3.3 & 1 & 1.3 & 9.0 & $2.75$ & $\pm$ & 0.15 & 1 & 97 & 17 & 40 & 1.92 \\
    HD 181327 & & 3.3 & 1 & 1.3 & 7.0 & $2.38$ & $\pm$ & 0.05 & 8 & 86 & 18 & 42 & 1.92 \\
    $\eta$ Crv & & 6.6 & 3 & 0.85 & 9.0 & $2.10$ & $\pm$ & 0.07 & 10 & 152 & 3 & 38 & 1.91 \\
    HD 95086 & & 8.6 & 1 & 1.3 & 7.0 & $2.41$ & $\pm$ & 0.12 & 1 & 208 & 19 & 35 & 1.91 \\
    $\beta$ Pic & & 8.7 & 1 & 0.87 & 7.0 & $2.81$ & $\pm$ & 0.10 & 8 & 106 & 20 & 49 & 1.93 \\
    HD 131835 & & 10.5 & 4 & 0.87 & 9.0 & $2.17$ & $\pm$ & 0.13 & 9 & 85 & 21 & 56 & 1.94 \\
    Fomalhaut & & 16 & 1 & 1.3 & 7.0 & $2.70$ & $\pm$ & 0.17 & 1 & 136 & 22 & 50 & 1.93 \\
    49 Cet & & 20 & 1 & 0.85 & 9.0 & $2.76$ & $\pm$ & 0.11 & 1 & 95 & 23 & 64 & 1.95 \\
    HR 4796\,A & & 27 & 5, 6 & 0.85 & 9.0 & $>2.73$ & $\pm$ & 0.10 & 9 & 78 & 24 & 73 & 1.96 \\
\hline
\end{tabular}
\tablebib{%
(1) \citet{macgregor+2016a}, (2) \citet{marshall+2018}, (3) \citet{marino+2017a}, (4) \citet{hung+2015a}, (5) \citet{gerbaldi+1999}, (6) \citet[parallaxes from GAIA Data Release 2]{gaia+2016,gaia+2018b}, (7) \citet{macgregor+2015b}, (8) \citet{ricci+2015b}, (9) \citet{marshall+2017}, (10) \citet{marino+2017a}, (11) \citet{matthews+2015}, (12) \citet{booth+2017}, (13) \citet{macgregor+2018b}, (14) \citet{marino+2018}, (15) \citet{liseau+2010}, (16) \citet{steele+2016}, (17) \citet{macgregor+2019}, (18) \citet{marino+2016}, (19) \citet{su+2017}, (20) \citet{matra+2019b}, (21) \citet{hung+2015b}, (22) \citet{macgregor+2017}, (23) \citet{hughes+2017}, (24) \citet{kennedy+2018a}.
  }
\end{table*}

\begin{figure}
  \centering
  \includegraphics{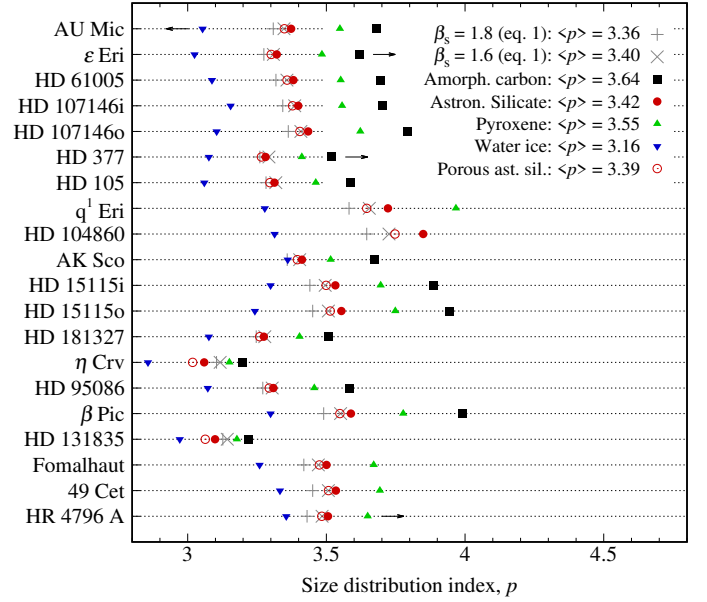}
  \caption{Size distribution indexes as derived from observed spectral indices for the discs presented in Table~\ref{tab:alpha-data}. Grey plusses and crosses mark the results obtained with Eq.~(\ref{eq:draine}) for $\beta\sbs{s} = 1.8$ and $\beta\sbs{s} = 1.6$, respectively. Black squares, filled and open red circles, green upward and blue downward triangles indicate results obtained with numerical Mie modelling for amorphous carbon, compact and porous Astronomical Silicate, pyroxene, and amorphous water ice, respectively. Where symbols are missing, the observed $\alpha\sbs{mm}$ could not be reproduced within the range $2.6 < p < 4.2$.\label{fig:macgregor}}
\end{figure}

Figure~\ref{fig:macgregor} shows the size distribution indexes resulting for a set of debris discs with observed $\alpha\sbs{mm}$ from Eq.~(\ref{eq:draine}) and from full Mie modelling, ordered by stellar luminosity. Table~\ref{tab:alpha-data} lists the values used for $\alpha\sbs{mm}$, $L$, disc radius $r$, wavelengths $\lambda_1$ and $\lambda_2$, together with the references. Temperatures, $T,$ and black-body spectral indexes, $\alpha\sbs{Pl}$ , are computed for mm-sized grains of astronomical silicate. This comparison shows three main effects.

First, the spread in size distribution indexes for a set of discs is wider for the more accurate numerical model than for Eq.~(\ref{eq:draine}). Because typically $1 < \beta\sbs{eff} < \beta\sbs{s}$, Eq.~(\ref{eq:draine}) tends to underestimate the sensitivity of $p$ on $\alpha\sbs{mm}$, while Eq.~(\ref{eq:alternative}) overestimates it. The numerical model falls in between. Extreme values near to or beyond the boundaries of the range $3 < p < 4$ are affected even more because there the $p(\alpha\sbs{mm})$ relation deviates more strongly from linearity. The wider spread not only can make outliers more significant but also any potential trend, such as $p(L_*)$ or $p(r)$.

Second, there is a systematic offset towards higher $p$ values in some of the new results. The lines defined by Eq.~(\ref{eq:draine}) and the numerical models typically cross around $p = 3.1 \ldots 3.3$ for the two compact silicates. With a mean $\langle p \rangle = 3.36 \pm 0.03$ for Eq.~(\ref{eq:draine}) and $\beta\sbs{s} = 1.8$ (or $\langle p \rangle = 3.41 \pm 0.03$ for $\beta\sbs{s} = 1.6$), most values are already higher than that. Hence, the increased spread also slightly increases the mean to $\langle p \rangle = 3.42 \pm 0.04$ for astronomical silicate and $\langle p \rangle = 3.57 \pm 0.05$ for pyroxene. For the porous silicate and the given set of discs, the mean (3.39) value is very close to that obtained for the simple $\beta\sbs{eff} = \beta\sbs{s} = 1.6$ relation.

Third, uncertainties in the composition of the dust cause uncertainties in both the mean of and the scatter among $p$ values. When pyroxene is assumed instead of astronomical silicate, for example, $\langle p\rangle$ is higher by 0.15 because the slope of the imaginary part of the refractive index of pyroxene is shallower.

\section{Discussion}\label{sec:Discussion}
Most of the results from the previous sections concern the discrepancies between a full numerical model and an analytic approximation based on power laws. The approximations leading to Eq.~\ref{eq:draine} tend to underestimate the mean slope and the spread of slopes of the grain size distributions for given observed spectral indexes. Assuming compact grains, the true size distribution slopes are somewhat steeper on average, which has implications for the processes that shape the size distribution. Values for $p > 3.5$ can be explained with details of the collisional cascade, such as specific material strength $Q\sbs{D}^*$ decreasing and/or collision velocities increasing with grain size \citep{o'brien+greenberg2003,pan+schlichting2012,gaspar+2012}. Values for $p < 3.5$ require a relative reduction of the lifetime of smaller grains. Poynting--Robertson or stellar wind drag would have that flattening effect on the size distribution \citep{reidemeister+2011,wyatt+2011}. The same is true for erosion by sublimation or UV sputtering \cite[e.g.][]{grigorieva+2007b,potapov+2018}, and for a stronger dynamical excitation of small grains by radiation pressure \citep{thebault+wu2008}. A correction towards higher values of $p$ could thus mean that collisions are more important, the according timescales shorter than that of drag and erosion.

A steeper size distribution would automatically imply a reduced overall disc mass when extrapolating from dust to planetesimals. If $p$ were higher by 0.1, the mass ratio between km-sized objects and mm-sized dust would be reduced by a factor of 4. This could be one of several aspects that help reduce the severity of the problem with the unrealistically high extrapolated disc masses that are derived for most of the more luminous debris discs \citep[e.g.][]{krivov+2018}. 

Additional uncertainties arise from a poor understanding of the dust composition and, hence, of its optical properties. On the one hand, the real part of the refractive index, $n$, determines the slope $1/\beta\sbs{eff}$ of the $p(\alpha\sbs{mm})$ dependence by defining the absorption and emission behaviour around the transition from transparent to opaque. For materials with higher $n$, such as non-porous amorphous carbon or silicates,
absorption of mid-sized grains is higher. The transition from small to opaque is more abrupt, corresponding to the two characteristic radii, $s\sbs{c}'$ and $s\sbs{c}$, being closer to one another. The transparent regime is more distinguished from the small-grain regime, but narrower. The coefficient $\beta\sbs{eff}$ is closer to unity. For materials with $n$ closer to unity, such as water ice or porous silicate, the transparent intermediate regime is less pronounced, but wider, and hence $\beta\sbs{eff}$ closer to $\beta\sbs{s}$, the power-law index of the volume absorption coefficient $A$. The fidelity of Eq.~\ref{eq:draine} thus depends on the material. Closer to the stars, where high-$n$ silicates are expected to dominate, the deviation that comes with approximation (\ref{eq:draine}) should be higher. Far from the stars, where low-$n$ volatile ices can exist in abundance, the deviation should be lower. Figure~\ref{fig:all-alpha-p} shows that the parameter region where Eq.~\ref{eq:draine} (with constants $\beta\sbs{s}$ and $\alpha\sbs{Pl}$) is fully applicable is, however, very small.

On the other hand, $\beta\sbs{s}$ is itself a material property and thus depends directly on, for example, the silicate-to-ice ratio. Measurement and usage of better optical data in the mm regime is required. Some of the commonly used data sets have never been designed to be used at such long wavelengths. The ice data set that was derived from a model fit to the $\beta$ Pictoris disc \citep{li+greenberg1998}, for example, inherits the uncertainties in the assumed size distribution. Figure~\ref{fig:k} shows that the adopted $k \propto \lambda^{-0.5}$ is flatter than the slope actually measured for ice at a few tens of K, leading to an overestimation of the size distribution index $p$. Temperature dependence in optical data at mm wavelengths has been discussed for some time \citep[e.g.][]{mishima+1983} but only recent observations at higher resolution and sensitivity have brought the issue back into focus. Consequently, temperature-dependent data for more and more materials are becoming available \citep{reinert+2015,haessner+2018,mutschke+mohr2019}.

Strong uncertainties remain even if a full Mie model is used and the grain composition is known. Two of the very basic assumptions, that of solid, spherical grains and that of a power-law size distribution, can be too crude. As shown in Sect.~\ref{sec:porosity}, the transition from absorption by cross-section to absorption by volume is smoother for filling factors $f \lesssim 0.5$, which is more in line with the assumptions that lead to Eq.~(\ref{eq:draine}). The lower bulk density and the lower refractive index, however, cause smaller grains and the lower size cutoff imposed by radiation pressure to become more important. The resulting relation between observed $\alpha\sbs{mm}$ and deduced $p$ flattens (corresponding to a higher $\beta\sbs{eff}$) for lower disc temperatures and stellar luminosities. If a constant $\beta\sbs{eff} = \beta\sbs{s}$ were imposed for porous grains, the size distribution index would be underestimated for less luminous stars or distant discs and overestimated for more luminous stars or closer discs, causing an artificial trend in disc statistics. If, indeed, porosity is significant, all results based on Mie theory of compact spheres end up suffering from a systematic error. \citet[][their Fig.~3]{brunngraeber+2017} show that the size distribution index for a filling factor of $f = 0.5,$ derived based on the assumption of solid grains, is already too high by 10 to 15\,\%, or, in absolute terms, 0.35 to 0.5 higher than their reference value of 3.5. However, their curves do not converge for even higher porosities (lower filling factors), indicating that there is no simple porosity threshold above which Eq.~(\ref{eq:draine}) is applicable and below which it is not.

Dust size distributions may deviate strongly from power laws when radiation pressure blowout introduces a discontinuity in the collisional cascade, resulting in a wavy pattern \citep{campo-bagatin+1994a,durda+dermott1997,thebault+2003,krivov+2006}. Parts of that waviness are carried over to the total opacity and the observable SED \citep{thebault+augereau2007}. The narrower the range of grain sizes that effectively contribute to emission at any given wavelength, the more of the waviness is preserved. This means that a wavier, more expository SED is produced by
the more abrupt transition from surface absorption to volume absorption that is seen for materials of higher optical density. For materials or wavelengths with a lower opacity, the convolution over a wider range of grain sizes reduces the waviness. In addition, in dynamically cold debris discs, smaller grains are under-abundant because they suffer from dynamical excitation due to radiation pressure \citep{thebault+wu2008}. If the transition from affected small to unaffected big grains occurs in the (sub)mm size range, the average slope derived from the SED analysis will be shallower than the background power-law slope of an infinite, unperturbed collisional cascade.
%  If porosity is low, the limiting grain size for radiation pressure blowout to occur is higher blowout grains of a few microns in size are much smaller than the important (sub-)mm grains. The waviness is reduced. However, if porosity is high, more notable waves may be imprinted in the mm spectrum.

Models ought to take into account that the characteristic grain size $s\sbs{c}$ can easily exceed $\lambda$ by orders of magnitude \citep[cf.][]{draine2006}: $s\sbs{c} \approx \lambda/(16k)$ for $n \approx 1$ and $s\sbs{c}' < s\sbs{c}$ in Eq.~\ref{eq:sc}. Assuming a moderate $k\sim 10^{-4}$ for water ice at $\lambda$ close to 1~cm, $s\sbs{c}$ is already in the order of $600 \lambda$, that is, several metres. For the more opaque compact carbon and silicates, where $k \gtrsim 0.01$ (Fig.~\ref{fig:k}), the result is: $s\sbs{c} \lesssim 6\lambda$. A porous blend of these material can be expected to fall in between these extremes. With the SED being dominated by grains with radii around $s\sbs{c}$ for $3 < p < 4$, the maximum grain sizes $s\sbs{max}$ in numerical and analytic models should always be set such that $s\sbs{max} \gg s\sbs{c}$ or reflect a physically motivated maximum. If the assumed $s\sbs{max}$ is too small, say 1~mm, a lack of flux results at the longer wavelengths. The resulting synthetic SED will be too steep, requiring a flatter-than-actual size distribution when fit to observational data. On the same note, any dust mass derived from mm or cm observations is a mass in cm or dm-sized grains (if these are present), which sets the starting point from which to extrapolate towards a total disc mass. The often quoted mass in mm grains and the total disc mass would then be lower by a factor of $(0.01\ldots 0.1)^{p - 3} \approx 0.1\ldots 0.3$ for $p=3.5$ \citep[cf.][]{krivov+2018}.

The large $s\sbs{c}$ and $s\sbs{max}$ imply that observations in the mm-cm wavelength range are sensitive to much bigger, dm-sized grains. Any derived size distribution index would also cover that size range, including potential changes in slope or cutoffs. Models for planetesimal formation through pebble aggregation, such as the streaming instability scenario \citep[e.\,g.][]{johansen+2015}, predict a transition from stronger individual pebbles to weaker and, hence, less abundant pebble piles around cm sizes \citep[cf.][]{krivov+2018}. This transition would translate to a steepening of the SED in the wavelength range of interest. Below and above the transition size from pebbles to piles, where collision physics are assumed continuous, the size distribution would be flatter. At earlier phases, when the pebbles are still forming from smaller grains, $s\sbs{max}$ is a measure of the actual maximum grain size. The opacity curves and observed SEDs are steeper early on, flattening as $s\sbs{max}$ grows with time \citep[e.\,g.][]{kruegel+siebenmorgen1994,natta+testi2004,draine2006}. At wavelengths for which optical depths $\gtrsim 1$ are reached, Eq.~(\ref{eq:basic}) is not applicable and, thus, the relations for $p(\alpha\sbs{mm})$ derived in Sects.~\ref{sec:Approximations} and \ref{sec:Numerics} are invalidated.

\section{Summary}\label{sec:Conclusions}
% Numerical and analytic computation of the thermal emission from circumstellar dust shows that the relation between dust grain size distributions and observed spectral slopes at long wavelengths is determined by the optical behaviour of grains in an intermediate size range where the transition from volumetric to surface absorption and emission occurs. This transition occurs more abruptly for materials with a higher refractive index, materials that have a higher optical density, that is. As a result, the slope of the SED becomes less sensitive to the material-dependent spectral slope of the imaginary part of the refractive index. Compact silicates and carbon-rich materials fall into that category. For these materials, the analytic approximation in Eq.~(\ref{eq:draine}) tends to underestimate the derived size distribution index, $p$, and how strongly it depends on the observed spectral index, $\alpha\sbs{mm}$. If the refractive index is closer to unity, the transition is smoother and the material dependence more direct. Hence, for less absorbing materials such as pure water ice, crystalline silicates, or porous materials, Eq.~(\ref{eq:draine}) improves in terms of its accuracy. No single linear relation $p(\alpha)$ is valid for all stellar luminosities and disc radii, however.

When analysing SEDs of individual systems, a full numerical approach is the most flexible in terms of materials and assumed size distributions. The computational cost of Mie calculations is low. If only a quick estimate of the size distribution index and its uncertainties is sought, Eq.~(\ref{eq:alpha-p}) can be used in combination with the coefficients provided in Fig.~\ref{fig:all-alpha-p}. These coefficients are fitted to the numerical results for moderate size distributions with $3.2 < p < 3.8$ for five materials, two wavelength combinations, and a range of stellar luminosities and disc temperatures.

With the exception of the unrealistic pure water ice, the numerical fits to observed systems result in steeper size distribution slopes, on average, tipping the scales more towards size-dependent material strength or impact velocities and, then, away from drag forces. The system-to-system variations are somewhat more pronounced than previously thought. However, these findings are accompanied by strong uncertainties related to the material composition and porosity. Further work is needed to examine the impact of realistic material blends, grain structures, and temperature-dependent optical properties at long wavelengths.

%------------------------------------------------------------------

\begin{acknowledgements}
TL is grateful to H. Mutschke for fruitful discussions on dust optical properties. Parts of this work have been supported by the \textit{Deutsche Forschungsgemeinschaft} (grants LO 1715/2-1 and MU 1164/9-1 within Research Unit FOR 2285 ``Debris Disks in Planetary System'').
\end{acknowledgements}

%  \bibliography{References}
%  \bibliographystyle{aa}

\end{document}